\documentclass[letter]{aa}
\usepackage{txfonts}
\usepackage{graphicx}
\usepackage{natbib}

\begin{document}

\title{Direct diameter measurement of a star filling its Roche Lobe}
\subtitle{The semi-detached binary SS Leporis spatially resolved with VINCI/VLTI}

\author{T. Verhoelst \inst{ \star}
  \and E. van Aarle
  \and B. Acke \thanks{Postdoctoral Fellows of
the Fund for Scientific Research, Flanders}}

\offprints{T. Verhoelst, \email{tijl.verhoelst@ster.kuleuven.be}}

\institute{Instituut voor Sterrenkunde, KULeuven, Belgium}

\date{Received 10 May 2007 / Accepted 26 May 2007}

\abstract{Stellar evolution in close binary systems is strongly
  influenced by mass transfer from one star to the other when one
  component fills its zero-velocity surface or Roche Lobe. SS\,Lep is a
  fairly nearby close binary showing the
  Algol paradox and a shell spectrum, both indicative of (past) mass
  transfer.}  {To study the process of mass transfer and its
  evolutionary consequences, we aim at a direct characterisation of the
  spatial dimensions of the different components of SS Lep with IR
  interferometry.}  {We use
  VINCI/VLTI interferometric observations in the K band and
  photometric observations from the UV to the far-IR. The visibilities
  are interpreted with simple geometrical models and the Spectral
  Energy Distribution (SED) is
  decomposed into the three main components: A star, M star and dust
  shell/disk.} {From the SED, we find that the main emitters in the K
  band are the M star and the circumstellar environment. Both are
  spatially resolved with the VINCI observations, showing the excess
  to be circumbinary and showing the M star to have a size equal to
  its Roche Lobe. } {We conclude that we have,  for the first time,
  directly resolved a star filling its Roche Lobe. The resulting mass
  transfer is probably the cause of (1) the circumbinary dust disk of
  which we see the hot inner region spatially resolved in our
  observations, (2) the unusually high luminosity of the A\,star and
  (3) the shell spectrum seen in the  UV and optical spectra.}


%\titlerunning{}
%\authorrunning{}

\maketitle

\section{Introduction}

Close binary evolution differs from single star evolution in many
respects. The most influential events after the main sequence (MS) are
undoubtedly the mass-transfer episodes. These alter the evolution
of both components, their chemical composition, their orbit and they
often lead to spectacular high-energy phenomena. Needless to say, such
mass transfer is a very complex event with many remaining questions
regarding, amongst others, stability, mass conservation and flow shape.

In her seminal paper, \cite{Cowley1967} suggests SS\,Leporis to be a
 semi-detached binary, consisting of a MS star and a giant which fills
 its Roche Lobe. The system shows the Algol paradox, i.e. the most
 evolved component appears the least massive, which indicates past
 mass transfer. SS\,Lep is not strictly an Algol binary as it shows no
 eclipses. This is a  consequence of its rather small orbital
 inclination. 

Before we discuss what is known about SS\,Lep to date, a preliminary
note on the distance to this object is essential since the debate
between Post- or Pre-MS classification has been vivid
\citep[e.g. ][]{Welty1995}. \cite{Polidan1991} mention an early
trigonometric parallax measurement of $30\pm14$\,mas which places
SS\,Lep as close as $33^{+30}_{-10}$\,pc. However, the {\sc hipparcos}
parallax is only $3.05\pm0.67$\,mas, corresponding to a distance of
$330^{+90}_{-60}$\,pc. Pourbaix (private comm.) confirms that any
effect due to binary motion is within the noise on the {\sc hipparcos}
measurements and that therefore the larger distance is the correct
one.

The binary nature of SS\,Leporis was first detected by
\cite{Wilson1914} in Lick Observatory spectrograms. It was found to
have a variable absorption spectrum by \cite{Struve1930} who also
showed that the velocity variations of the sharp metallic lines are
inconsistent with binary motion but rather represent phenomena
produced in an expanding atmosphere. \cite{Smith1942} also found
redward-displaced emission edges at the low Balmer members and at the
strongest \ion{Fe}{ii} and \ion{Ti}{ii} lines. \cite{Slettebak1950}
detected TiO lines in the IR spectrum of SS\,Lep indicating the
presence of an M-type companion.  \cite{Widing1966} confirmed the
presence of this M companion and proposed a Roche Lobe overflow model
developed by \cite{Kraft1958} for T\,CrB as working hypothesis.

\cite{Cowley1967} was the first to determine the spectroscopic orbit
(of the M star) accurately, see Table\,\ref{tab:orbit}. Assuming an ad
hoc radius for the M\,star of 75\,R$_{\odot}$ and an inclination of
$24\degr$, she found that the M\,star fills its Roche Lobe but only at
periastron. The orbital parameters were refined by \cite{Welty1995}
with a significant change in excentricity: $0.024 \pm 0.005$ instead
of 0.132. They concluded that the mass-transfer at/after periastron
scenario of \cite{Cowley1967} is therefore unlikely.

\begin{table}
\caption{Relevant orbital parameters. $P$ is the period, $e$ the
  excentricity, $K$ the velocity amplitude of the M\,star, $a$ the
  semi-major axis of the M\,star orbit, $i$ the inclination and
  $f(M)=(M_{\rm{A}}\sin{i})^3/(M_{\rm{A}}+M_{\rm{M}})^2$.} 
\label{tab:orbit}
\centering
\begin{tabular}{lcc}
\hline\hline
                 & \cite{Cowley1967} & \cite{Welty1995} \\
\hline
$P$ [days]       & $260\pm0.3$       & $260.34\pm1.80$ \\
$e$              & $0.132\pm0.043$   & $0.024\pm0.005$ \\
$K$ [km s$^{-1}$]& $21.0\pm0.8$      & $21.32\pm0.21$  \\             
$a\sin{i}$ [10$^6$\,km] & 74         & $76.30\pm0.53$  \\
$f(M)$           & 0.24              & $0.261\pm0.005$ \\
\hline
\end{tabular}
\end{table}

Mid-IR emission features of silicate dust in the spectrum of SS\,Lep
were first identified by \cite{Allen1972} and \cite{Jura2001} suggested
that SS\,Lep is surrounded by a circumbinary dust disk with large
grains, responsible for the far-IR and mm fluxes, from which an
evaporation wind consisting of smaller particles arises and generates
the excess at wavelengths below 100\,$\mu$m. 

We report here on the direct size determination of the M star and the
circumbinary matter using near-IR interferometry.

\section{Observations}

\subsection{Photometry}
\label{sec:obs:phot}

The photometric observations used to construct the SED were taken from
the literature and span the entire wavelength range from the UV to the
far-IR. Geneva 7-colour observations were taken from
\citet{Rufener1976}. Near-IR magnitudes are those from the 2MASS PSC
\citep{Skrutskie2006} and the far-IR magnitudes were found in the IRAS
PSC \citep{Neugebauer1984}. Additional observations in H,K and L are
available from the CIT survey of early-type emission-line stars
presented by \cite{Allen1973}. Variability in the optical is very
small: 0.013\,mag with a period of 130\,days, i.e. exactly
$P_{\rm{orb}}/2$ \citep{Koen2002}. Variability in K is also very small
with a detection at the $3\sigma$\,level, $\Delta K = 0.15$ by
\cite{Kamath1999}.

\subsection{Interferometry}

Interferometric observations in the near-IR (K band) were obtained
with the VLTI commissioning instrument VINCI on baselines with lengths
ranging from 4.7 to 121\,m and position angles (PAs) from 54 to
85\,$\degr$ East-of-North. The 338 observations cover the period from
12 December 2002 to 30 December 2003. Instrumental visibilities were
derived with the standard data reduction pipeline, v3.1
\citep{Kervella2004}. The visibility calibration was done by linear
interpolation of the interferometric efficiency throughout the night.
The stars used as calibrators are: $\nu2$\,CMa (K1III, 2.38\,mas),
31\,Ori (K5III, 3.56\,mas), 51\,Ori (K1III, 1.87\,mas), HR\,2311
(K5III, 2.43\,mas), HR\,2549 (K5III, 2.19\,mas) and HR\,2305 (K3III,
1.76\,mas). The Uniform Disk (UD) diameters were taken from
\citet{Borde2002}.

\section{Modelling}

In order to have some a priori knowledge on the angular sizes and the flux
contributions of the different components in the K band, we
first construct an SED. 

\subsection{The SED}

We use the photometry presented in Sect.\,\ref{sec:obs:phot} to
construct the SED and fit\footnote{The fitting is done by comparing
observed integrated flux over the filter transmission profile with the
same quantity computed from the synthetic SEDs.} a combination of a
Kurucz model for the A star and a {\sc marcs} model for the M star to
the Geneva and 2MASS J,H observations.  We searched for the best
reproduction of the observations among the models (solar metallicity,
spaced 250\,K in $T_{\rm{eff}}$ and 0.5\,dex in $\log{g}$) having
temperatures and surface gravities roughly compatible with the
spectral types \citep[A1 and between M3 and M4.5
respectively,][]{Welty1995}. We chose to exclude the K and L band
observations in our fit since the SED presented by \cite{Jura2001} and
the VINCI observations (see Sect.\,\ref{sec:modelling:vis}) indicate
some circumstellar excess already at these wavelengths. We do not fit
the mid- to far-IR excess either since it is dominated by cold dust
emission \citep[450\,K, ][]{Fajardo1995}. The remaining free variables
are the two stellar diameters and the total amount of extinction,
i.e. both interstellar and circumstellar. The wavelength dependence of
the extinction is modelled using the law of \citet{Cardelli1989} with
R$_V = 3.1$.

We find T$_{\rm{eff}} = 9000\pm250$\,K and log\,$g=1.5\pm0.5$ for the A\,star and
T$_{\rm{eff}} = 3250\pm250$\,K and log\,$g=1.0\pm0.5$ for the M\,star.  We note
that the Geneva colours agree much better with such a low surface
gravity for the A\,star than with that of a MS star, which is
independent evidence that the A\,star is indeed located above the
MS. A surface gravity of log\,$g$=1.48 was already derived from the
Geneva colours by \cite{Hauck2000} but they neglected the influence of
the companion. The M\,star's spectral type is rather uncertain, but we
find a strong constraint on its temperature from our SED: the J and
H\, band observations sample the peak of the intensity distribution
(the H$^-$ opacity minimum) in an M\,star and they are therefore very
sensitive to its temperature. Our temperature is slightly below the
temperature of a typical M4III giant and corresponds more to spectral
type M6III \citep{Perrin1998}.

\begin{figure}
 \resizebox{\hsize}{!}{\includegraphics{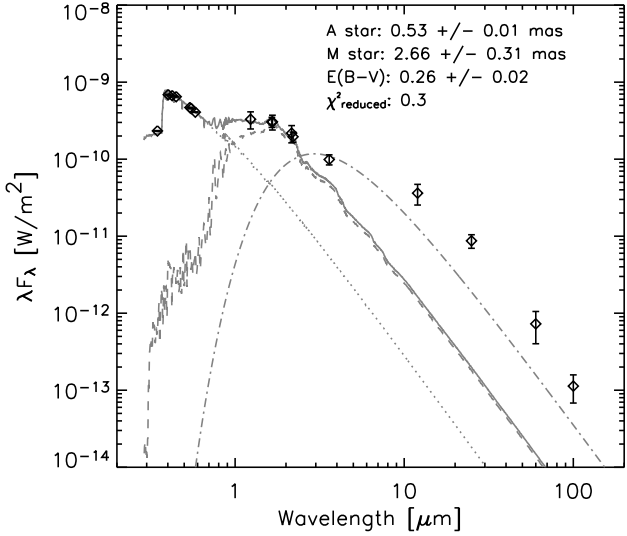}}
 \caption{SED, consisting of Geneva, 2MASS, CIT and IRAS photometry,
 together with our model (solid line) consisting of the A and M stars
 (dotted and dashed lines respectively). Also shown is the blackbody
 curve at the brightness temperature (1250\,K) and diameter (10.5\,mas
 FWHM) of the circumbinary dust disk generating the near-IR excess, as
 determined from the interferometric observations (see
 Sect.~\ref{sec:theexcess}). To notice the excess in the K\,band, one
 should realize that the bandpass actually covers the CO bandhead seen
 on the right of the datapoint.}
 \label{fig:sed}
\end{figure}

Our preferred model ($\chi^2_r = 0.3$) and its parameters are shown in
Fig.\,\ref{fig:sed}. We find clear evidence for a near-IR excess in
both the K and L band observations. In the K\,band, at the wavelength
of our interferometric observations, the A\,star contributes
$11\pm7$\% of the flux, the M\,star roughly 60\%. The remainder is
unaccounted for in our model.

\subsection{The K band visibilities}
\label{sec:modelling:vis}
\begin{figure}
 \resizebox{\hsize}{!}{\includegraphics{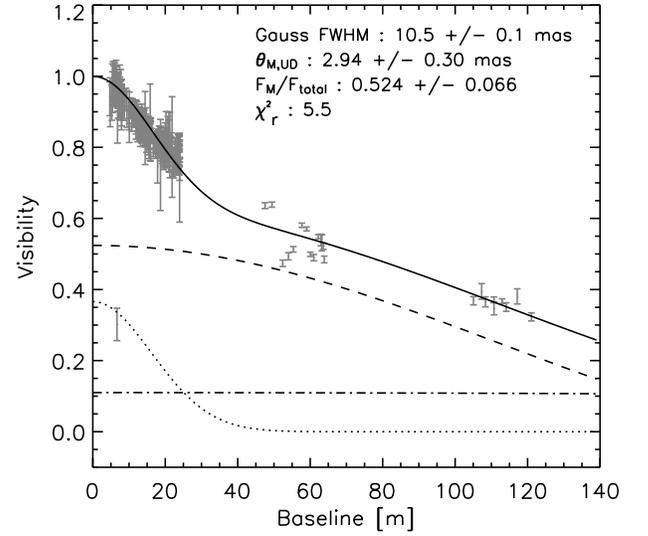}}
 \caption{VINCI/VLTI K band visibilities and our model (solid line)
 consisting of a Gaussian Disk and 2
 Uniform Disks. The contribution to the total visibility of the
 extended component (Gaussian), the M\,star (UD) and the A\,star (UD)
 are shown with a dotted, dashed and dash-dot line respectively. Note
 that the A\,star diameter and flux and the uncertainties on these
 values were taken from the SED fit.}
 \label{fig:visfit}
\end{figure}

Several geometrical models are confronted with the data: a single UD,
two and three concentric Uniform and/or Gaussian Disks and a binary
model with possibly resolved components. A single UD yields a $\chi^2_r =
39$, and is thus clearly incompatible with the observations.  A fit
with two concentric UDs or a large Gaussian Disk and a smaller Uniform
Disk, having three free parameters: the 2 diameters and the flux
ratio, has a $\chi^2_r = 6.4\, \rm{or}\, 5.5$ respectively. We should
also include the unresolved contribution of the A\,star. The baselines
are not long enough to constrain flux contribution and diameter of the
A\,star so we use the parameters\footnote{We computed the conversion factor from the
  Limb-Darkened (LD) diameter derived in SED fitting to the UD
  diameter used for the interferometric modelling from our atmosphere
  models and found it to be 0.945 in the K band for the M\,star
  parameters and 0.970 for the A\,star.} derived from the SED\,fit:
$\theta_{\rm{A,UD}} =
0.51\,$mas and F$_A$/F$_{total} = 0.11 \pm 0.07$ in K. This model,
shown in Fig.\,\ref{fig:visfit}, agrees rather well with the
observations ($\chi^2_r = 5.5$) and adding complexity, i.e. a possible
offset between components, results in a degenerate problem: the
$\chi^2_r$ hypersurface does not constrain the additional parameters
(length and angle of the separation vector), and shows an unreasonable
minimal $\chi^2_r = 0.07$. Fortunately, the derived diameters appear
unsensitive to the introduction of such an offset. We must conclude
that, while we can trust the derived diameters, the current dataset
does not allow for a characterisation of possible offsets (which are time-dependent since the orbital period is of the
order of 1\,yr), and this is therefore the aim of future AMBER and MIDI
observations (see Sect.\,\ref{sec:conclusions}).  The inclusion of the
barely resolved A\,star does not change the quality of the fit, but it
does have a small influence on the derived diameter for the M star. We
find the total uncertainty on the M\,star diameter and flux
contribution with the inclusion of the uncertainty in A\,star flux to
be 0.27\,mas (10\,\%) and 7\% respectively.

\section{Discussion}
\label{sec:discussion}

We summarize the results of our SED model and interferometric
observations in Table\,\ref{tab:results}. The flux ratios and
diameters used in our geometrical model
make an association with the components seen in the SED
straightforward: the UD of
2.8\,mas is the M\,star and the Gaussian Disk is the extended component
generating the excess emission in K.  

\begin{table}
\caption{Stellar and CSE parameters derived from the SED and
  interferometric observations. NA indicates that this quantity could
  not be (reliably) measured, ``idem'' indicates that the SED-derived
  value was used.}
\label{tab:results}
\centering
\begin{tabular}{lcc}
\hline\hline
                                & SED            & VINCI \\
\hline
$T_{\rm{eff,A}}$ [K], $\log{g} $ & 9000, 2.0           & NA    \\
$T_{\rm{eff,M}}$ [K], $\log{g} $ & 3250, 1.0           & NA    \\
$\theta_{\rm{A,LD}}$ [mas]         & $0.53\pm0.02$  & idem \\
$\theta_{\rm{M, LD}}$ [mas]         & $2.66\pm0.33$  & $3.11\pm0.32$ \\
%$\theta_{\rm{cold~dust}}$ [mas] & $48.6\pm26.1$  & NA \\
$\theta_{\rm{hot~excess}}$  [mas] & NA             & $10.5\pm0.1$ \\
$L_{\rm{A}}$ [L$_{\odot}$]      & $1900\pm250$   & NA    \\
$L_{\rm{M}}$ [L$_{\odot}$]      & $1200\pm400$   & NA  \\
$F_{\rm{A}}$ at 2.2~$\mu$m       & $11\pm7\%$    & idem \\
$F_{\rm{M}}$ at 2.2~$\mu$m       & $60\pm50\%$   & $52\pm7\%$ \\
$F_{\rm{hot~excess}}$ at 2.2~$\mu$m & NA    & $38\pm7\%$  \\
\hline
\end{tabular}
\end{table}

\subsection{The M\,star: filling its Roche Lobe}

Converting the measured UD diameter of the M\,star to a physical LD
diameter yields $\theta_{\rm{M,LD}}=3.11\pm0.32$\,mas or
R$_{\rm{M}}=110\pm30 \rm{R}_{\odot}$. The uncertainty on the linear
diameter includes that on the distance. This is in good agreement with
the LD diameter estimated from the SED fit. Given the absence of
strong variability and the O-rich nature, we believe it to be on the
Red Giant Branch (RGB). 

The constraint on the A\,star mass used by \cite{Welty1995} to derive
the inclination should be relaxed somewhat towards the upper end since
we have now the confirmation that it is of luminosity class II. This
yields an inclination of $30\degr\pm10\degr$.  Using $(a_{\rm{A}} +
a_{\rm{M}})\sin{i}=98.1\pm2.9$\,million km or $a = (141 \pm 4
R_{\odot})/\sin{i}$, and the generally assumed mass ratio
$M_{\rm{A}}/M_{\rm{M}}=1/q = 4\pm1$, we find
%for the M\,star a
%Hill Sphere Radius $r_{\rm{H}} =  a(\frac{m}{3M})^{1/3} = 113\pm20
%\rm{R}_{\odot}$. The Hill Sphere clearly has the same size as the
a Roche radius\footnote{$R_{\rm{Roche}}=a \frac{0.49 q^{2/3}}{0.6
  q^{2/3}+\rm{ln}(1+q^{1/3})}$, \citep{Eggleton1983}}
  $R_{\rm{Roche,M}} =
%  a \times 0.27\pm0.02 = 
74^{+40}_{-20}\rm{R}_{\odot}$, where the errorbar includes the
  $10\degr$ uncertainty on the inclination.

From the observed surface gravity and radius of the M\,star, and
the use of a reasonable mass for the A\,star (see
Sect.\,\ref{sect:astar}), we find that the mass ratio
$M_{\rm{A}}/M_{\rm{M}}$ could in fact be a little lower. This would
reduce the size of the Roche Lobe for the M\,star. 

The find that the M\,star radius is equal to the
Roche radius, which is very strong evidence that the M\,star is at
this moment completely filling its critical Roche equipotential and
therefore that mass transfer must be taking place. This is confirmed
by the  shell spectrum with significant UV activity
\citep{Polidan1994}, the presence of a circumbinary dust disk and the
excess luminosity of the A\,star as shown below. 

\subsection{The excess: a circumbinary dust disk}
\label{sec:theexcess}
The extended component we see in the interferometric observations has
a size clearly larger than the binary separation and we are therefore
certain that almost half of the K\,band flux comes from a circumbinary
structure.  We determine its brightness temperature to be
$T_{\rm{B}}=1250$\,K. This temperature is in agreement with the fact
that we see the excess only from the near-IR onward and it corresponds
more or less to the sublimation temperature of oxygen-based dust
grains. This suggests that we are observing the hot inner region of
the dusty circumbinary structure also generating the mid- to far-IR
excess seen in the SED. A clue to differentiate between spherical or
disk-like geometry is the amount of reddening. We find
$A_V=0.7\pm0.1$\,mag in our SED fit, which is in good agreement with
the $E(B-V)=0.26$ derived by \cite{Malfait1998} based on IUE
data. This $A_V$ corresponds to
$\tau_{\rm{dust}} \sim 0.5$ at visual wavelengths. Since the
line-of-sight extinction is either due to interstellar dust grains or
due to circumstellar grains outside the orbital plane, we can assume them
to be small\footnote{A
  spherical dust shell which is stable enough to allow significant
  grain growth is implausible: the radiation pressure would quickly
  drive them away.}.  We can therefore use the optical
constants of \cite{Dorschner1995} for amorphous silicates to
extrapolate the $\tau_{0.55\mu\rm{m}}$ to IR wavelengths. We find an
optical depth of only $\tau_{2.2\mu\rm{m}} = 0.03$ in the K band. The
optical depth required to generate the K\,band excess at temperatures
below that for dust sublimation is of the order of unity, and we can
thus conclude that the dust geometry is disk-like rather than
spherical, and seen almost face-on, as expected from the system's
orbital inclination.  Note that a temperature much higher than the
derived $T_{\rm{B}}$ would generate excess emission also at shorter
wavelengths, which is not seen in the SED. We can therefore rule-out
the presence of a hot spherical circumbinary shell with significant
continuum opacity. 

\cite{Jura2001} suggest a circumbinary dust disk with an evaporation
wind  to explain the mid en far-IR spectral properties of the
excess. In their scenario the disk contains mainly large ($d \sim
0.1$\,mm) grains. These would have an optically thin temperature at our derived
distance from the central stars of 1480\.K which is in rough
agreement with our derived value of 1250\,K. The difference is
probably due to the disk being fairly optically thick. The disk wind,
consisting of smaller grains, is possibly responsible for the minimal
amount of extinction still observed towards the central stars.

The presence of this dusty disk shows that the mass transfer is not
conservative which is in agreement with theoretical predictions that
the additional radiation pressure in the cool M\,star atmosphere may
modify the critical Roche equipotential into a surface including both
the inner and outer Lagrangian points \citep{Schuerman1972}.

\subsection{The A star: accreting}
\label{sect:astar}

The reaction of the primary, which has a radiative envelope, to the
accretion should be expansion, as its adiabatic exponent is positive
\citep{Ritter1996}.  Indeed, we find a radius roughly ten times as
large as that of a typical A1V star (R$_{\rm{A}}\sim
18$\,R$_{\odot}$), while its mass, derived from the observed surface
gravity and diameter, is not that of a (super)giant: $0.4
\rm{M}_{\odot} < M_{\rm{A}} < 4\rm{M}_{\odot}$. \cite{Kippenhahn1977}
determined evolutionary tracks for mass-accreting MS stars and find
that the excursion from the MS mass-luminosity relation depends on the
accretion rate and initial mass. The position of the A\,star of
SS\,Lep in the HR diagram lies on the track of a 2\,M$_{\odot}$ star
accreting $2\times10^{-4}$\,M$_{\odot}$/yr, after roughly
1\,M$_{\odot}$ has been accreted. For the mass ratio to become as
large as is  generally assumed ($M_{\rm{A}}/M_{\rm{M}} = 4$), this requires that
the M\,star has also lost a substantial amount of mass through the
circumstellar disk or a wind: otherwise, the M\,star would have been
the least massive to begin with, which is incompatible with its more
evolved status.

\section{Conclusions and outlook}
\label{sec:conclusions}

We conclude that SS\,Lep is a semi-detached binary, consisting of a
late-M giant which fills its Roche Lobe and therefore loses mass to
the primary star. The latter was most probably a regular MS star but now
shows an increased size and luminosity due to a relatively high accretion
rate. The mass transfer appears to be non-conservative since we detect
an optically thick circumbinary dust disk. The shell spectrum detected
at UV and optical wavelengths, in general not expected in A\,type
stars, is thus most likely due to this ongoing mass transfer. We
believe this to be the first time a Roche-Lobe filling star and the
resulting circumbinary disk have been directly resolved using
optical/IR interferometry.

We have planned further interferometric observations with AMBER and
MIDI on the VLTI to characterize in more detail the shape and
structure of both the M\,star and the circumstellar excess. We have
also started a new spectroscopic campaign with the CORALIE instrument
on the Swiss telescope at La~Silla, Chile to improve the determination
of the orbit and to study the gas flow dynamics.  
 
Algol-type binaries such as SS\,Lep can provide a
 wealth of information on mass transfer and binary
 evolution. Optical/IR interferometric observations such as those
 presented here, are well suited for studies of low-inclination
 systems and in that way complementary to eclipse and doppler-imaging
 work.

\acknowledgements{The authors would like to thank the anonymous
  referee for useful comments. Based on observations made with ESO's VLT
  Interferometer at Cerro Paranal, Chile. The VINCI data were
  retrieved from the ESO/ST-ECF Archive.}

\bibliographystyle{aa}
\bibliography{references}

\end{document}